\documentclass[onecolumn,aps,superscriptaddress,nofootinbib]{revtex4}

\usepackage{graphicx,amsmath}

\newcommand{\beq}{\begin{equation}}
\newcommand{\eeq}{\end{equation}}
\newcommand{\bea}{\begin{eqnarray}}
\newcommand{\eea}{\end{eqnarray}}

\newcommand{\Eq}[1]{(\ref{#1})}

\newcommand{\one}{\ensuremath{\mathbf{1}}}

\newcommand{\mo}{\ensuremath{\mathcal{O}}}
\newcommand{\go}{\ensuremath{\gamma_\mathcal{O}}}
\newcommand{\ignore}[1]{}

\makeatletter
\def\simgt{\mathrel{\lower2.5pt\vbox{\lineskip=0pt\baselineskip=0pt
           \hbox{$>$}\hbox{$\sim$}}}}
\def\simlt{\mathrel{\lower2.5pt\vbox{\lineskip=0pt\baselineskip=0pt
           \hbox{$<$}\hbox{$\sim$}}}}
\makeatother

\begin{document}
\setlength{\unitlength}{1mm}

\title{Higgs mass naturalness and scale invariance in the UV}

\author{Gustavo Marques Tavares}
\author{Martin Schmaltz}
\affiliation{Physics Department, Boston University, Boston, MA 02215}
\author{Witold Skiba}
\affiliation{Physics Department, Yale University, New Haven, CT 06520}
\affiliation{CERN Physics Department, Theory Division, CH-1211, Geneva 23, Switzerland}

\begin{abstract}

It has been suggested that electroweak symmetry breaking in the Standard Model may be natural if the Standard Model merges into a conformal field theory (CFT) at short distances. In such a scenario the Higgs mass would be protected from quantum corrections by the scale invariance of the CFT\@. In order for the Standard Model to merge into a CFT at least one new ultraviolet  (UV) scale is required at which the couplings turn over from their usual Standard Model running to the fixed point behavior. We argue that the Higgs mass is sensitive to such a turn-over scale even if there are no associated massive particles and the scale arises purely from dimensional transmutation. We demonstrate this sensitivity to the turnover scale explicitly in toy models. Thus if scale invariance is responsible for Higgs mass naturalness, then the transition to CFT dynamics must occur near the TeV scale with observable consequences at colliders. In addition, the UV fixed point theory in such a scenario must be interacting because logarithmic running near a free fixed point constitutes hard breaking of scale invariance and spoils the Higgs mass protection.

\end{abstract}

\maketitle

\section{Introduction}

The observation \cite{Higgsdiscovery} of a light Higgs boson with properties which are consistent with the Standard Model (SM) has motivated much reexamination of the notion of naturalness \cite{naturalness}. In theories of natural electroweak symmetry breaking the Standard Model is modified at high scales to incorporate a symmetry which protects the mass of the Higgs boson from sensitivity to high scales. Supersymmetry or the shift symmetries of Little Higgs theories require the introduction of  ``partner" particles which cancel the contributions to the Higgs mass from Standard Model particles. The cancellation between SM particles and partners occurs for each value of the loop momentum. 

A qualitatively different proposal is to use conformal symmetry to protect the Higgs mass.\footnote{Scale invariance is sufficient to protect the Higgs mass. It is believed that for unitary and causal $4d$ quantum field theories scale and Lorentz invariance imply conformal invariance \cite{Luty:2012ww,Fortin:2012hn}, therefore we do not distinguish between scale and conformal invariance in this paper.}
The idea is that if the Higgs scalar field were part of a conformal field theory (CFT), then its mass would be forbidden by scale invariance. Of course, the SM has particle masses and running couplings which break conformal symmetry. However, if conformal symmetry could be broken sufficiently ``softly" such that the symmetry is restored at high energies, then the Higgs mass would still be protected from the largest radiative corrections which come from the highest energies. 

The idea to use scale invariance to protect the Higgs mass was proposed long time ago in \cite{bardeen} and has recently received some attention \cite{scaleinvariance}. In the recent literature, scale invariance of the classical Lagrangian is often invoked to argue the absence of large quantum corrections. This is misguided because only symmetries of the quantum theory constrain the form of possible counter terms. Thus to employ scale invariance for protecting the Higgs mass one has to show that all SM coupling constants approach fixed points when evolved into the ultraviolet (UV) with the renormalization group.\footnote{A less ambitious proposal is to merge the non-gravitational couplings into a CFT between the weak and the Planck scale~\cite{frampton} and assume that unspecified quantum gravitational dynamics sets the desired conformal boundary conditions at Planck scale. Our results apply to either case.} An obvious problem is that in the SM both gravitational and hypercharge couplings grow at short distances. If unchecked, each coupling becomes non-perturbatively large at a characteristic scale in the UV, leading to very large conformal symmetry breaking near the Planck scale and the hypercharge Landau pole, respectively. Thus, at the least hypercharge and gravitational interactions must transition from their current evolution to the CFT behavior before the couplings become non-perturbatively large. If this transition involves ultra-heavy particles these particles can appear in virtual corrections, and it is well-known that loops of massive particles destabilize the Higgs mass. But what if there are no heavy particles at the transition scales? And are there constraints on the evolution of asymptotically free couplings like the QCD coupling which do not grow in the UV?

In this paper we show that the Higgs mass is sensitive to any threshold scales in the UV, including scales which only arise from dimensional transmutation and are therefore hidden from Feynman diagram calculations at fixed loop order. Our argument relies on the fact that anomalous dimensions of operators which couple to the Higgs change as the threshold is crossed. Given this sensitivity of the Higgs mass to the threshold scale, it is clear that naturalness requires the transitions to happen at low scales (near the TeV scale for hypercharge and below $\sim(M_{\rm Weak} M_{\rm Planck})^{1/2}$ for gravity).

The proposal of a CFT in the UV requires us to think carefully about the notion of fine tuning. In the usual effective field theory picture we envision a cutoff beyond which the laws of physics are unknown (generally at or below the Planck scale), and counter terms at the cutoff encode contributions from unknown physics above the cutoff. A theory is then natural if the coefficients of operators in the effective theory below the cutoff are stable against quantum corrections in the effective theory. 

It is important to distinguish between technically natural parameters like a chiral fermion mass which are protected by symmetries and parameters which are not protected by symmetries. For the chiral fermion, all contributions to the mass are proportional to chiral symmetry breaking. Therefore, the renormalized mass can be kept fixed by scaling chiral symmetry breaking to zero at the same time as taking the regulator to infinity. In contrast, a parameter which is not protected by symmetries receives additive contributions from renormalization. If these contributions are divergent then fine-tuning of the bare parameter against quantum corrections is required to keep the parameter finite as the regulator is taken to infinity. While such tuning is a mathematical possibility, it is unnatural unless it is enforced by a symmetry.  Scale invariance can play the role of this symmetry, but since scale invariance is broken in the Standard Model one must ensure that scale invariance violation is soft, i.e.  no new divergences are introduced by the couplings which break scale invariance. In the following sections, we  explain this criterion with specific examples and show
that the approach to the UV fixed point has to be sufficiently rapid so that conformal symmetry breaking does not enter the Higgs mass corrections from the UV\@. Only then is it natural to invoke UV scale invariance to cancel away the Higgs mass contributions from the UV\@. We find that deviations of the gauge couplings from their UV fixed point values must scale away as sufficiently large powers of distance at short distances. For example, the logarithmic approach to the free fixed point of the QCD coupling in the SM is too slow, and contributions to the Higgs mass due to the scale invariance violation from to the running QCD coupling persist to arbitrarily high energies. Thus, even the asymptotically free couplings of the SM must transition to different UV behaviors for the Higgs mass to be protected by scale invariance.

Given our results, it is clear that a successful implementation of ``Higgs mass naturalness from scale invariance in the UV" requires major modifications of the Standard Model at the TeV scale. In particular, all gauge couplings must turn over and approach non-trivial fixed points near the TeV scale because free fixed points always lead to logarithmic running. We emphasize that our results apply irrespective of the choice of regulator. Dimensional regularization is often most convenient for calculations, but the apparent absence of power divergences in dimensional regularization does not remove sensitivity to high scales and associated fine tuning.  

In Section \ref{sec:scale} we review the calculation of quantum corrections to the Higgs mass in perturbative theories and discuss the connection to scale invariance. In Section \ref{sec:examples}  we define toy models for field theories which transition between different fixed point behaviors in the UV and IR\@. We show explicitly that the Higgs mass is sensitive to the threshold scale in the anomalous dimensions which result from the differing UV and IR scalings. 
In Section \ref{sec:af} we show that dimensionless couplings which are asymptotically free lead to logarithmic breaking of scale invariance in anomalous dimensions. Hence in asymptotically free theories, scale invariance cannot protect the Higgs mass from large corrections in the UV\@. We conclude in Section \ref{sec:conclusions}.

\section{Higgs mass sensitivity to the UV and scale invariance}
\label{sec:scale}

\begin{figure}[ht]
\includegraphics[width=0.2\textwidth]{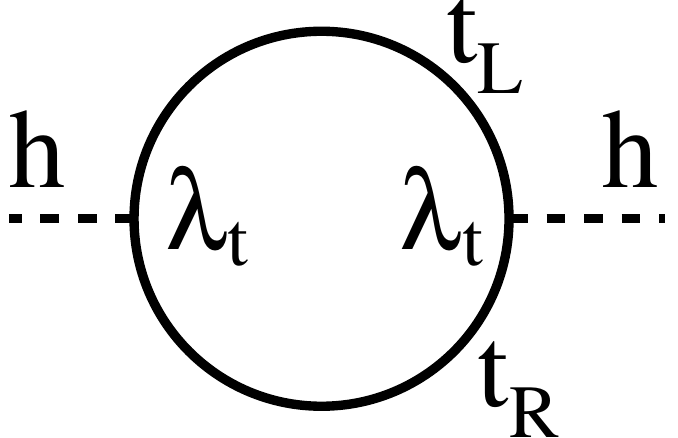}
\caption{Top quark loop contribution to the Higgs mass.}
\label{fig:toploop}
\end{figure}

Consider a toy SM with a Dirac fermion ``top quark" $t$ with a chiral coupling to a complex scalar ``Higgs"
\beq
\mathcal{L} = -  \lambda_t (h \, \overline t P_L t + h^\dagger \, \overline t P_R t )\ .
\label{eq:lagrangian}
\eeq
Assuming a non-vanishing fermion mass,\footnote{Or equivalently expanding about non-zero $h$ expectation value. We also introduce the renormalization scale $\mu$ to account for the change in dimension of the coupling $\lambda_t$ in $d$ dimensions. The scale is arbitrary and will not be very important for our discussion.} $m_t$, the usual top quark contribution to the Higgs mass (Figure \ref{fig:toploop}) is 

\bea
\delta m_h^2 &=& - i N_c \lambda_t^2 \mu^{4-d} \int \frac{d^dp}{(2\pi)^d} Tr
 \left[P_L \frac{\slash\!\!\!{p}+m_t}{p^2-m_t^2+i \epsilon} P_R  \frac{\slash\!\!\!{p}+m_t}{p^2-m_t^2+i\epsilon} \right] \nonumber \\
&=& -2 i N_c \lambda_t^2 \mu^{4-d} \int\frac{d^dp}{(2\pi)^d} \frac{p^2}{(p^2-m_t^2+i\epsilon)^2} \nonumber \\
&=& -  \frac{N_c \lambda_t^2}{8 \pi^2} m_t^2 \left(\frac{4 \pi \mu^2}{m_t^2}\right)^\frac{4-d}{2}
\frac{d}{2} \Gamma(1-\frac{d}{2}) =  
\frac{N_c \lambda_t^2}{4 \pi^2} m_t^2\left(\frac1{\epsilon}-\gamma-\log\frac{m_t^2}{4\pi\mu^2}+\frac12\right)
\eea
near $d=4-2\epsilon$. This becomes $\delta m_h^2 = -  m_t^2 N_c \lambda_t^2/(4 \pi^2) (\log[m_t^2/\mu^2]-\frac12) $ after modified minimal subtraction $\overline{\rm MS}$ of the UV divergence. Famously, there is no quadratic divergence and the correction is proportional to a finite scale, the top quark mass $m_t$. 

However, the answer is regulator dependent. In fact, there are additive contributions from any finite momentum shell above $m_t$ which grow with momentum squared. If we had used a momentum cutoff $\Lambda$ these contributions would have been more explicit, and we would have had to remove the quadratic divergence ``by hand" to obtain a finite answer. In absence of a symmetry, the remaining Higgs mass is arbitrary because the counter term which cancels the divergence is only determined up to an additive constant. And since the contributions from short distances grow without bound, any finite Higgs mass requires fine tuning the counter term.  Dimensional regularization with $\overline{\rm MS}$ is misleading in this case because it does the tuning automatically, but it does not change the fact that there is sensitivity to high scales and additional finite pieces could be added to the counter terms. 

To decide whether the fine-tuning in this example is intrinsic to the theory or just an artifact of a poorly chosen regulator requires careful consideration of the symmetries. In particular, we could imagine that our top-Higgs system is embedded in a larger theory which is conformal in the UV\@. The conformal symmetry is broken by the top mass, but since the mass is a relevant operator one might try to argue that conformal symmetry is restored in the UV\@. If that were true, a vanishing Higgs mass in the UV would correspond to an enhanced symmetry point. Then a Higgs mass counter term which cancels all contributions from the top loop above $m_t$ would be perfectly natural. Unfortunately, this argument is not correct and even this simple example is more subtle.

To understand the subtlety,  consider first the same calculation as above with a vanishing top mass. Then our toy theory really is conformal and the above argument applies. In dimensional regularization we would compute
\beq
\delta m_h^2 \propto \int \frac{d^dp}{(2\pi)^d} \frac{1}{p^2} = 0 \ .
\eeq
The vanishing result is a consequence of scale invariance of the theory. The same result  could have also been obtained with a momentum cutoff $\Lambda$ as a regulator. This regulator clearly violates conformal invariance, and the integral would have been proportional to $\Lambda^2$. However, the underlying scale invariance of the theory requires us to choose the counter term such as to cancel this contribution exactly.

We are now ready to consider the effects of conformal symmetry breaking from the top mass. This adds a new logarithmic divergence proportional to $m_t^2 \log( \Lambda^2/m_t^2)$. The explicit $m_t$-dependence of the divergence makes it manifest that conformal invariance is not restored in the UV\@. There contributions to the Higgs mass grow without bound as the cutoff is taken to infinity, and canceling them with a counter term to obtain a finite Higgs mass constitutes fine-tuning. In particular, for exponentially large cutoffs $\Lambda$ there is no sense in which a Higgs mass near the top quark mass is preferred by the symmetries.

How does this compare with the innocuous looking result obtained in dimensional regularization with $\overline{\rm MS}$? The dimensional regularization answer before inclusion of the counter term does contain the divergence in the form of the $1/\epsilon$ term. After subtraction, the renormalized mass is finite because the contributions from the UV have already been fine-tuned away. We can still see them in the dependence of $\delta m_h^2$ on $\mu$. Taking $\mu$ exponentially larger than $m_t$ we see that the running $\overline{\rm MS}$ Higgs mass at large $\mu$ becomes much larger than $m_t$. The small value at $\mu \sim m_t$ is a result of fine-tuning the UV Higgs mass parameter in order to obtain a small mass in the IR\@.

In the following, we will find it convenient to calculate the Higgs mass contribution in position space
\beq
\delta m_h^2 = i (-i\lambda_t)^2 \mu^{4-d} \int d^d x
\langle 0| T\, \mathcal O^\dagger (x) \mathcal O(0) | 0\rangle .
\label{eq:Higgsmass}
\eeq
This reproduces the momentum space result after plugging in the (normal ordered) operator $\overline t P_L t $ for $\mathcal{O}$, contracting fields with propagators, and integrating over $x$
\bea
&&  \int d^d x \
\langle 0| T \, (\overline t P_R t)(x)\, (\overline t P_L t)(0) | 0\rangle  \nonumber \\
&=&-N_c \int d^d x \,
\, Tr \left[ \int \frac{d^dp}{(2\pi)^d} e^{ipx} P_L 
\frac{i(\slash\!\!\!{p}+m_t)}{p^2-m_t^2+i \epsilon} P_R  
\int \frac{d^dq}{(2\pi)^d} e^{-iqx}\frac{i(\slash\!\!\!{q}+m_t)}{q^2-m_t^2+i\epsilon} \right] \nonumber \\
&=& 2 N_c  \int \frac{d^dp}{(2\pi)^d} \frac{p^2}{(p^2-m_t^2+i\epsilon)^2} \ .
\eea

How would the position space computation be modified if the the Higgs coupled to an interacting CFT? 
To have an example theory to consider, we construct a class of perturbative models similar to those proposed by Banks and Zaks \cite{BZ}. Consider QCD with a large number of colors $N_c$ and the number of flavors chosen such that the gauge coupling has a weakly-coupled IR-stable fixed point. Perturb the theory by coupling a single complex scalar ``Higgs" with a flavor-universal Yukawa coupling to all quarks of the CFT as in \Eq{eq:lagrangian}. The coupling is a relevant perturbation because $\mo=\overline t P_L t$ has negative anomalous dimension. The one-loop beta function for the Yukawa coupling also has an IR stable fixed point. Setting the gauge and Yukawa couplings to their fixed point values $g=g^*$ and $\lambda_t=\lambda_t^*$ in the UV, this theory is conformal at all scales.\footnote{In addition, a quartic coupling for the scalar is generated from fermion loops. The quartic coupling also has a non-trivial fixed point. The contributions from the quartic coupling at the fixed point are suppressed by powers of $1/N_c$ and we ignore them for simplicity.} This defines an interacting CFT with a negative anomalous dimension \go\ for $\mo$.

The two point function of \mo\ in  \Eq{eq:Higgsmass} is determined by conformal invariance
\beq
\langle 0| T \mathcal O^\dagger (x) \mathcal O(0) | 0\rangle =
C \left(\frac{1}{-x^2+i\epsilon}\right)^{d-1+\gamma_{\mathcal O}},
\label{eq:twopointcft}
\eeq
where $x^2=t^2-\vec x^2$. The constant $C$ depends on the normalization of the operator which we choose such that
$C=1$.
Now the computation of the Higgs mass in the CFT is simple. To lowest order in $\lambda_t$, we obtain 
\beq
\delta m_h^2 = -i \lambda_t^2 \mu^{4-d} 
\int d^d x \left(\frac{1}{-x^2+i\epsilon}\right)^{d-1+\gamma_\mathcal O} = 0 
\eeq
in dimensional regularization. As in the case of free field theory, dimensional regularization automatically discards contributions to the Higgs mass from the cutoff. Of course, we could have also computed this with any other regulator, and we would have determined the Higgs mass counter term to restore scale invariance.

\section{Higgs mass sensitivity to non-perturbative threshold scales}
\label{sec:examples}

We now turn to the case of interest: a theory which transitions between two different scaling behaviors at a transition scale $M$. Since the interacting fixed point for the couplings $g$ and $\lambda$ of the theory we introduced in the previous section is IR attractive, one could use that theory to construct an example of a model which flows from a free fixed point in the UV to the interacting one in the IR with an associated change in dimension of the operator \mo. However, we are interested in a theory which is IR free and flows to an interacting UV fixed point because this more similar to what one would expect if the hypercharge $U(1)$ was to merge into a CFT in the UV\@.

In such a theory, the two-point function \Eq{eq:twopointcft} is of the form
\beq
\label{eq:2ptwithf}
  \langle 0| T\,  \mathcal O^\dagger (x) \mathcal O(0) | 0\rangle = \left(\frac{1}{-x^2}\right)^{d-1} f(-x^2M^2) \ .
\eeq
Here, the factor of $(-x^2)^{1-d}$ is determined by dimensional analysis and it coincides with  the two-point function in free field theory. The dimensionless function $f(y)$ contains all the information about the interacting dynamics. Since $f(y)$ is defined over all scales it cannot be computed reliably at any fixed order in perturbation theory. At the very least, calculating $f(y)$ requires knowing the solution to the renormalization group equations to determine anomalous dimensions.  In a conformal regime, $f(y)$ reproduces the power law corresponding to the anomalous dimension of ${\mathcal O}$. In the transition region between two fixed points,  it interpolates between the two power laws appropriate for the fixed points in the UV and IR\footnote{Near the free fixed point in the IR the scaling of operators includes logs which we ignore. We will discuss logs near a UV fixed point in the next Section.}
\beq
  f(y)\rightarrow\left\{ \begin{array}{lll} 1 & {\rm as}& y\rightarrow \infty \  (\mathrm{IR}) ,\\  
y^{-\gamma_{\mathrm{UV}}} & {\rm as}& y \rightarrow 0 \ \ \, (\mathrm{UV}).  \end{array}  \right.
\eeq

For our purposes, the most important feature of the two-point function in \Eq{eq:2ptwithf} is that it depends on the transition scale $M$ in a non-trivial way. This $M$-dependence is what ensures that the $\int\! d^d x$ integral in \Eq{eq:2ptwithf} does not vanish even in dimensional regularization barring fine tuning\footnote{An exception arises when the $M$-dependence is trivial and can be factored out of the integral. Then the integrand does not transition between two different scaling behaviors in the UV and IR, and the corresponding integrals, $\int d^d x (-x^2)^\alpha$, can be set to $0$ in dimensional regularization.}. This is the main result of our paper: the Higgs mass does receive quantum corrections proportional to the non-perturbative transition scale $M$.

To see this more explicitly, we simplify \Eq{eq:2ptwithf} by performing a Wick rotation, using the $SO(d)$ symmetry of Euclidean space to do the angular integrals, and changing variables to $y=-x^2 M^2$. Then
\beq
\delta m_h^2 = - i\lambda_t^2 \mu^{4-d} \int d^d x \left(\frac{1}{-x^2}\right)^{d-1} f(-x^2M^2) = 
  - M^2\, \frac{\lambda_t^2 \pi^{d/2}}{\Gamma(d/2)}\left(\frac{\mu^2}{M^2}\right)^{2-d/2}
  \int_0^\infty \frac{dy}{y^{d/2}} \,  f(y)\ .
\label{eq:Higgsmassf}
\eeq

We now consider two illustrative examples for $f(y)$ and compute the Higgs mass in each case. We chose our examples based on calculability and their qualitative features. The first example is a crude toy model that is far from realistic, but easy to understand analytically. We assume that the transition between the UV and IR fixed points is abrupt
\beq
  f(y)=\left\{ \begin{array}{lll} 1 & {\rm for}& y>1, \\   y^{-\gamma_{\mathrm{UV}}} & {\rm for}& 1>y>0.  \end{array}  \right.
\label{eq:crude1}
\eeq
The integral in \Eq{eq:Higgsmassf} is divergent in the UV (at small $y$) and requires a regulator. However since the UV behavior of the correlation function is exactly that of a CFT we can use scale invariance of the UV theory to fix the mass counter term to cancel the scale invariance violation from the regulator. A nice way to implement the subtraction by the counter term is to subtract from \Eq{eq:Higgsmassf} the corresponding expression in a CFT: $\int dy / y^{d/2}$. This automatically subtracts the correct counter term required by scale invariance of the UV theory in any regularization scheme. Of course, in the special case of dimensional regularization this counter term vanishes so that we simply subtract zero. If after the subtraction we are left with a finite integral then we can evaluate it in four dimensions. 

The integral from $0$ to $1$ is completely removed by the subtraction and the remainder gives
\beq
 \delta m_h^2 = -M^2 \pi^2\lambda_t^2  \int_{1}^\infty \, \frac{dy}{y^2} \left[1-\frac{1}{y^{ \gamma_{\mathrm{UV}}}}  \right] =
 - M^2 \pi^2 \lambda_t^2 \frac{\gamma_{\mathrm{UV}}}{ 1+ \gamma_{\mathrm{UV}}} .
\eeq
In this case, the main contribution to the Higgs mass can be thought of as coming from the threshold at $M$ (or $y=1$) where there is an abrupt change of the scaling dimension. 

\ignore{We next consider a more gradual transition between the UV and IR by including a transition region. Now there are three different power law scalings in three regions: the IR, intermediate, and UV
\bea
  f(y)=\left\{ \begin{array}{lll} 
   1 & {\rm for}& y>c, \\  
  (y/c)^{-\gamma_{\mathrm{UV}}/2}  & {\rm for}& c>y> 1/c,\\  
   y^{-\gamma_{\mathrm{UV}}} & {\rm for}& 1/c>y>0\  .\end{array}  \right.
\eea
For $c\rightarrow 1$ this reduces to the previous example. The UV divergence is unchanged and we make the same subtraction as before. Hence
\bea
 \delta m_h^2 &=&-M^2 \pi^2\lambda_t^2 \left\{ \int_{1/c}^{c}  \, \frac{dy}{y^2} \left[\left(\frac{c}{y}\right)^{\gamma_{\mathrm{UV}}/2}\!\!\!\!\!\!\!- \frac{1}{y^{\gamma_\mathrm{UV}}} \right] +\int_{c}^\infty \, \frac{dy}{y^2} \left[1-\frac{1}{y^{ \gamma_{\mathrm{UV}}}} \right] \right\}\nonumber \\
  &=&-M^2 \pi^2\lambda_t^2\, \frac{\gamma_{\mathrm{UV}}}{ (1+ \gamma_{\mathrm{UV}})(2+ \gamma_{\mathrm{UV}})} \left[ c^{1+\gamma_{\mathrm{UV}}}+ \frac{ 1+ \gamma_{\mathrm{UV}}}{c} \right].
 \eea

While the detailed expression is not particularly enlightening we point out that this reduces to the previous result for $c=1$ and grows for $c>1$. The growth with $c$ stems from the additional UV region near $y\sim1/c$ which dominates the integral. Thus even though the transition is smoother in this example, the fact that more of the UV contributes leads to a {\it larger} contribution to the Higgs mass.

For our last example, }

The abrupt transition in our example \Eq{eq:crude1} is clearly unphysical and one might think that a smoother and more physical transition would suppress the contributions to the Higgs mass. In fact, we will find the opposite to be true in our next example. We consider the smooth function $f(y)$
\beq
    f(y)=\left( \frac{1}{y^{n \gamma_\mathrm{UV}} + y^{n \gamma_\mathrm{IR}}} \right)^{1/n} \ .
    \label{eq:smooth}
\eeq
Assuming that $ \gamma_{\mathrm{UV}} < \gamma_{\mathrm{IR}}\leq 0$, $\gamma_{\mathrm{UV}}$ modifies the two-point function at small $y$, while $\gamma_{\mathrm{IR}}$ does so at large $y$. Meanwhile, the power $n$ controls whether the transition between UV and IR is smooth or abrupt. The larger the $n$ the more abrupt the transition.
\begin{figure}[ht]
\includegraphics[width=0.6\textwidth]{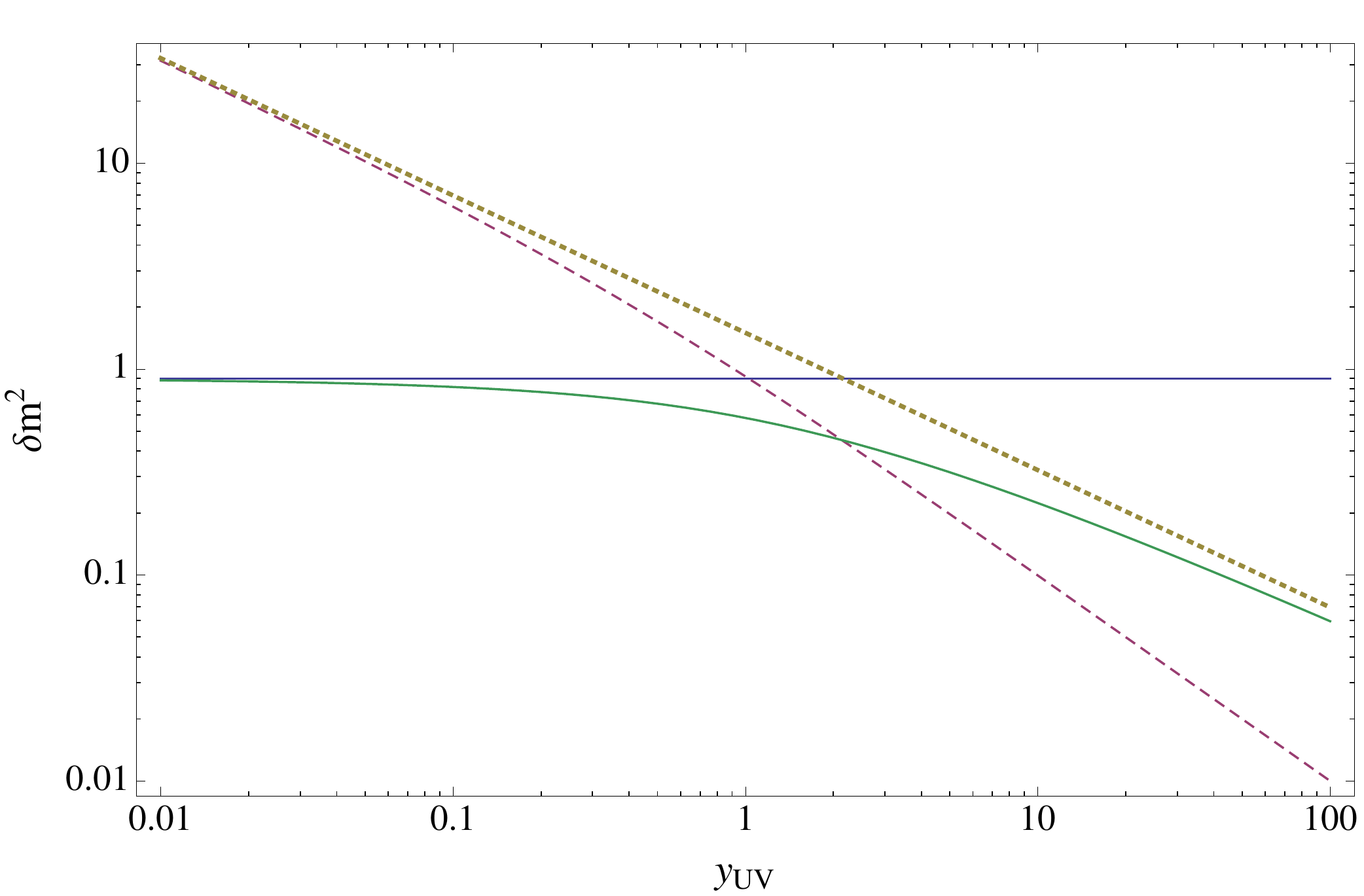}
\caption{Plot of the contributions to the Higgs mass in \Eq{eq:example3} as a function of the UV cutoff $y_\mathrm{UV}$ on the integral. The blue horizontal line corresponds to the full Higgs mass contribution. The green line which asymptotes to it at small $y_\mathrm{UV}$ is \Eq{eq:example3} as a function of the short distance cutoff $y_\mathrm{UV}$. The two lines which grow at small $y_\mathrm{UV}$ are the original unsubtracted integral (red-dashed) and the subtraction (yellow-dotted), both evaluated with the short distance cutoff. In this example, we chose $n=4$ for the transition parameter and $\gamma_\mathrm{UV}=-1/3$ for the UV anomalous dimension.}
\end{figure}

Choosing free field theory for the IR, $\gamma_\mathrm{IR}=0$, we have
\bea
 \delta m_h^2 &=&-M^2 \, \frac{\lambda_t^2 \pi^{d/2}}{\Gamma(d/2)}\left(\frac{\mu^2}{M^2}\right)^{2-d/2} \int_{0}^{\infty}  \, \frac{dy}{y^{d/2}}  \left( \frac{1}{1 + y^{n \gamma_\mathrm{UV}}} \right)^{1/n} \nonumber \\
&\xrightarrow{d=4}&-M^2 \, \lambda_t^2 \pi^2\int_{0}^{\infty}  \, \frac{dy}{y^2}  \left[\frac{1}{ \left( 1 + y^{n \gamma_\mathrm{UV}}\right)^{1/n}} -\frac{1}{y^{\gamma_\mathrm{UV}}}\right]
 = - M^2 \,\lambda^2_t \pi^2 \,  \frac{ \Gamma \left( \frac{1}{n} +\frac{1}{ n \gamma_{UV} } \right) \Gamma \left(1-\frac{1}{n \gamma_{UV}} \right) }{\Gamma \left( \frac{1}{n} \right)} \ ,
\label{eq:example3}
\eea
where in the second line we first subtracted a quantity which vanishes in dimensional regularization and then set $d=4$. The subtraction is identical to the contribution to the Higgs mass in a CFT which has identical UV anomalous dimensions to our example theory. It corresponds to the counter term which is required by scale invariance of the UV theory.
The remaining integral is finite for sufficiently large $n$ and $-1<\gamma_\mathrm{UV}<0$.

It is interesting to explore where the main contribution to the Higgs mass comes from. To this end we compute the subtracted integral in \Eq{eq:example3} with a UV cutoff $y_\mathrm{UV}$. The integral evaluates to  a generalized hypergeometric function. We plot it as a function of $y_\mathrm{UV}$ in Figure 3 for $n=4$ and $\gamma_\mathrm{UV}=-1/3$. One sees that the Higgs mass contribution approaches its full value near $y_\mathrm{UV}=1$ implying that the bulk of the Higgs mass contribution comes from quantum corrections with energies near the transition scale $M$. Where exactly the maximum of the contributions comes from depends on the parameters $n$ and $\gamma_\mathrm{UV}$.

What happens for smaller values of $n$ for which the subtracted integral is not UV finite? For small 
$n$ the transition between the IR and UV fixed point is relatively slow. As in the top quark loop example in  Section 2, conformal invariance is not restored sufficiently rapidly in the UV, and new UV divergences which depend on conformal symmetry breaking arise. 
To see these UV divergent contributions explicitly we expand the function $f(y)$ in \Eq{eq:example3} in a series for small $y$. The leading term is removed by the CFT subtraction. But for small $n$ one or several subleading terms are also UV divergent. These terms give contributions which are proportional to fractional powers of the UV cutoff and of $M$ and are evidence that scale invariance is not restored in the UV\@. Thus in these examples the Higgs mass suffers from sensitivity to the UV cutoff entangled with the transition scale $M$.

Can the Higgs mass be suppressed by considering smoother transitions? We have already seen that $n$ needs to be large enough to just obtain a finite contribution. Considering only values of $n$ such that the integral in \Eq{eq:example3} is finite, the magnitude of the Higgs mass contribution increases for smoother transitions (smaller $n$). This is because smoother transitions increase the width of the transitions region so that contributions from shorter distances can contribute in  \Eq{eq:example3}.

We close this Section with a comment on the choice of regulator. So far, we have mostly used dimensional regularization and relied on the fact that integrals of power laws, $\int d^d x (-x^2)^\alpha$, can be consistently set to 0~\cite{dimregreview}. How would our calculation differ if we used an explicit momentum cutoff $\Lambda$?

Considering for example the calculation of the Higgs mass in the smooth transition model we would find
\beq
\begin{split}
  \delta m_h^2 = -M^2 \lambda_t^2 \pi^2 \int_{\frac{M^2}{\Lambda^2}}^\infty \frac{dy}{y^2} f(y) 
  = -\lambda^2_t \pi^2 \left( \frac{\Lambda^2 }{1+\gamma_{UV} } \left( \frac{\Lambda^2}{M^2} \right)^{\gamma_{UV} } +  M^2 \, \frac{ \Gamma \left( \frac{1}{n} +\frac{1}{ n \gamma_{UV} } \right) \Gamma \left(1-\frac{1}{n \gamma_{UV}} \right) }{\Gamma \left( \frac{1}{n} \right)} + \dots \right) ,
\end{split}
\label{eq:lambdasmooth}
\eeq
where the dots stand for terms which vanish as we take $\Lambda$ to infinity. The second term reproduces the dimensional regularization result. The first term arises from the explicit breaking of scale invariance by the cutoff and its apparent $M$-dependence is a fake which stems from our choice of normalization of the operators \mo\ and $H$. In our normalization, the operators are normalized to their free (IR) scaling behavior. However, in the UV the operators have anomalous dimensions which modify the divergence to $\Lambda^{2+\gamma_{UV}}$, the explicit power of $M$ in \Eq{eq:lambdasmooth} arises from the matching of the UV Higgs mass operator to the  IR\@.  Note that the $M$-dependence of the subleading divergences for small $n$ is different, it corresponds to breaking of scale invariance that persists in the UV\@.

\section{Asymptotically free theories}
\label{sec:af}

The example theory considered in the previous section has an interacting fixed point in the UV\@. In the vicinity of an interacting fixed point correlation functions do not exactly obey the power-law scaling in \Eq{eq:twopointcft}, but the deviations from such scaling are suppressed by higher powers of $x^2$. We have already seen that if the approach to the UV fixed point is too slow, i.e. deviations from CFT scaling are not suppressed by sufficient powers of $x^2$ then the scale invariant UV subtraction is not sufficient to render the Higgs mass finite. In this case the theory requires fine-tuning the UV counter terms to cancel contributions which depend on both the UV and the threshold scale $M$.  

In the case of a free UV fixed point the approach to the free CFT scaling is especially slow. It includes logarithms which break conformal invariance at arbitrarily short distances. We therefore do not have a symmetry which allows canceling the UV contributions to the Higgs mass.  As a result it is not possible to disentangle Higgs mass sensitivity to the transition scale $M$ from sensitivity to the cutoff $\Lambda$.

To see the problem arise explicitly in a calculation consider again a complex scalar coupled to the ``top quark" as in \Eq{eq:lagrangian}. For simplicity,  assume that ``top quark" is massless to avoid the scale invariance breaking from the mass. We assume that the quark has its usual asymptotically free QCD interactions as in the Standard Model, and we are interested in computing the contributions to the Higgs mass from the UV, taking into account the running QCD coupling.  

The UV contributions to the matrix element \Eq{eq:twopointcft} can be computed using the operator product expansion. The coefficient of the leading $\one$ operator in $ \mathcal{O}^\dagger(x)\mathcal{O}(0)$ gives (see for example Chapter 18.3. in \cite{peskin})
\beq
 \langle 0| T\,  \mathcal O^\dagger (x) \mathcal O(0) | 0\rangle \propto \left(\frac{1}{|x|^2}\right)^{3}
\left(\log\frac{1}{|x|^2 M^2}\right)^{-a/b_0} \ .
\eeq
If we were discussing real QCD, $M$ would be replaced by $\Lambda_{\rm QCD}$, $a=8$ would be the one-loop anomalous dimension coefficient of ${\mathcal O}=\overline{t} t$, and $b_0=7$ for 6 quark flavors. The scale $M$ appears because we have re-summed an infinite class of diagrams to capture the leading $x$-dependence at small $x$ using the renormalization group.

We now integrate this expression over $x$, and change variables to $z=1/x^2$ to obtain
\beq
\delta m_h^2 \propto
  \int_{z_{IR}}^\infty dz \left(\frac{1}{\log(z/M^2)}\right)^{a/b_0} \ .
\eeq
We also introduced an IR cutoff somewhat above the transition scale $M$ (or $z_{IR} \sim {\rm few} \times M^2$) to exclude distance scales from the integration for which the running of the coupling deviates significantly from the one-loop approximation. Assuming that there are no further scales above $M$ one would naively expect this integral to be given by some order one factor times $M^2$. However, the integral is divergent in the UV and requires regularization. 

As discussed earlier, it is not apparent how to do this while keeping the contributions from $M$ and the regulator separate. Evaluating the integral in $d<2$ dimensions and analytically continuing to $d=4$ one encounters a multi-valued function with branch cuts in the complex $d$-plane, which makes the analytic continuation not unique. Subtracting the corresponding integral of the UV CFT  ($\int_{z_{IR}}^\infty \! dz \, 1$) to try to restore scale invariance fails because of the inverse logarithm. Evaluating the integral with a momentum cutoff one obtains terms which scale roughly like $\Lambda^2 $ times inverse powers of $\log(\Lambda^2/M^2)$ in the UV and therefore require $M$-dependent subtractions to remove the cutoff dependence.
In summary, it is not possible to separate cutoff dependence from $M$ dependence because scale invariance is not preserved in the UV\@. We conclude that theories with coupling constants which approach free fixed points in the UV cannot protect the Higgs mass from fine tuning. 

\section{Conclusions}
\label{sec:conclusions}

It has been proposed that if the SM were to merge into a conformal field theory in the UV then the scale invariance of the asymptotic CFT would guarantee the cancelation of UV contributions to the Higgs mass and the Higgs mass could be natural.  The proposal necessarily introduces new UV scales, the scales at which the couplings constants stop running as in the SM model and instead start approaching their UV fixed points.  
We showed that quantum corrections to the Higgs mass are sensitive to these scales even if there are no massive particles and the scales are of non-perturbative origin.

Therefore, naturalness of electroweak symmetry breaking in the SM requires that any such non-perturbative scales are sufficiently low. In particular, this means that the hypercharge beta function must be modified near the TeV scale and the running of the gravitational coupling must change at or below $\sqrt{M_\mathrm{weak} M_\mathrm{Planck}}$. 

Of course, the beta function of the hypercharge coupling can be computed easily in the Standard Model, and it does not have a UV fixed point. Thus, turning around the coupling at the TeV scale requires new interactions beyond the SM. In principle, gravitational interactions could provide these new contributions to the beta function, however at the TeV scale gravitational interactions are much too weak to be relevant. Hence, new interactions are required near the TeV scale, and it seems that unification of the $U(1)$ into a non-Abelian gauge group is the simplest possibility. Therefore, new weakly-interacting gauge bosons with masses near the TeV scale are expected in theories in which the SM is merged into a CFT in the UV\@. This is, of course, no different from more conventional approaches to the hierarchy problem in which new particles at the TeV scale are required to fill out multiplets of the symmetry which protects the Higgs mass from quantum corrections.

In addition to the dependence on threshold scales we also showed that anomalous dimensions must approach their UV fixed point values sufficiently rapidly for conformal symmetry of the UV to protect the Higgs mass from divergent contributions from short distances. This rules out even asymptotically free gauge groups as possible ingredients of a UV CFT\@.

\section*{Acknowledgments}
We thank Walter Goldberger, Ami Katz, David Poland, Ennio Salvioni, Javi Serra, and especially Andy Cohen for useful discussions which helped our understanding.
The work of MS and WS was partially supported by the US Department of Energy and GMT acknowledges support from a DOE High Energy Physics Graduate Fellowship.  We thank the Galileo Galilei Institute for Theoretical Physics where this project originated for the hospitality and the INFN for partial support.  WS thanks the CERN Theory Group for its hospitality and the Simons Foundation for their support. This work was partially supported by a grant from the Simons Foundation (\#267592 to Witold Skiba).


\end{document}